\documentclass{article}

\usepackage[utf8]{inputenc}

\usepackage[english]{babel}
\usepackage{subeqnarray}
\usepackage{url}

\usepackage{graphicx,psfrag}
\usepackage{fixltx2e,amsmath}
\usepackage{xcolor}
\usepackage{amssymb}
\usepackage[final]{pdfpages}
\usepackage{bbold}
\usepackage{dsfont}
\usepackage{pgfplots}
\usepackage{subfig}
\usepackage{algorithm}
\usepackage{algorithmic,placeins}
\usepackage[font=small,labelfont=bf]{caption}

\usepackage[font={footnotesize}]{caption}

\usepackage[utf8]{inputenc}
\usepackage[T1]{fontenc}
\usepackage{xcolor,amsmath,amssymb}
\usepackage{geometry}
\usepackage{graphicx}
\usepackage[sort&compress,numbers]{natbib}
\usepackage{enumerate}
\usepackage{wasysym}
\usepackage[textsize=footnotesize]{todonotes}

\usepackage{hyperref}
\hypersetup{
    colorlinks=true,
    linkcolor=red,
    filecolor=magenta,      
    urlcolor=cyan,
}

\usepackage{pgfplots}

\definecolor{darkblue}{rgb}{0,0,0.6}
\definecolor{darkred}{rgb}{0.6,0,0}

\usepackage{hyperref}

  \def\erf{\text{erf}}

\def\to{\rightarrow}

\newcommand{\beq}{\begin{equation}} \newcommand{\eeq}{\end{equation}}

\newcommand{\sgn}{\operatorname{sign}}
\newcommand\be{\begin{equation}}
\newcommand\bea{\begin{eqnarray} \nonumber }
\newcommand\ee{\end{equation}}
\newcommand\eea{\end{eqnarray}}

\usepackage{authblk}

\usepackage[bitstream-charter]{mathdesign}

\usetikzlibrary{pgfplots.groupplots}

\graphicspath{{figures/}} 

\title{The Inelastic Market Hypothesis:\\ A Microstructural Interpretation}
\author{Jean-Philippe Bouchaud \\
Capital Fund Management, 23 rue de l'Université, 75007 Paris \\ Chair of Econophysics and Complex Systems, \\ 
Ecole polytechnique, 91128 Palaiseau Cedex, France\\
Acad\'emie des Sciences, Quai de Conti, 75006 Paris
}

\date{January 2022}

\begin{document}

\maketitle
\abstract{We attempt to reconcile Gabaix and Koijen's (GK) recent Inelastic Market Hypothesis (IMH) with the order-driven view of markets that emerged within the microstructure literature in the past 20 years. We review the most salient empirical facts and arguments that give credence to the idea that market price fluctuations are mostly due to order flow, whether informed or non-informed. We show that the Latent Liquidity Theory of price impact makes a precise prediction for GK's multiplier $M$, which measures by how many dollars, on average, the market value of a company goes up if one buys one dollar worth of its stocks. Our central result is that $M$ is of order unity, as found by GK, and increases with the volatility of the stock and decreases with the fraction of the market cap. traded daily. We discuss several empirical results suggesting that the lion's share of volatility is due to trading activity. We argue that the IMH holds for all asset classes, beyond the case of stock markets considered by GK.
}

\section{Introduction}

Traditionally, market prices are considered to reflect the fundamental value (of a stock, currency, commodity, etc.), up to small and short-lived mispricings. In this way, a financial market is regarded as a measurement apparatus that aggregates all private estimates of an asset's true (but hidden) value and, after a quick and efficient digestion process, provides an output price. In this view, private beliefs should only evolve because of the release of a new piece of information that objectively changes the value of the asset. Prices are then martingales because (by definition) new information cannot be anticipated or predicted. In this context, neither microstructural effects nor the process of trading itself can affect prices, except perhaps on very short time scales.

This Platonian view of markets is fraught with a wide range of difficulties that have been the subject of thousands of academic papers in the last 40 years. The most well-known of these puzzles is the excess-volatility puzzle \cite{Shiller,Summers}: prices move around too much to be explained solely in terms of fluctuations of the fundamental value. But there is also the excess-trading puzzle \cite{Odean} and the trend-following puzzle (see e.g. \cite{Momentum1,Momentum2,Lemperiere} and refs. therein): Investors trade far too much and price returns tend to be positively autocorrelated on long times scales, such as weeks to months. In other words, some information about future price moves seems to be contained in the past price changes themselves. This is in stark contradiction with the efficient market story. 

Faced with these puzzles, research in the 1980s proposed to break away from the strict orthodoxy of rational market participants, and to instead introduce a new category of uninformed agents (or \textit{noise traders}) \cite{Black,noise1,Summers}. Including such noise traders allows one to account for excess trading. In fact, as illustrated by the Kyle \cite{Kyle} and Glosten--Milgrom \cite{Glosten-Milgrom} models, the existence of noise traders is crucial for liquidity providers to a least break even --- without them, liquidity would vanish and markets would not even function. However, in these models, prices still do not deviate from fundamentals: noise traders do participate in the price-formation process, but the impact of their trades does not contribute to price volatility \cite{Kyle} -- only ``true'' information can change prices. 

After accepting the presence of non-rational agents, the next conceptual step is to accept that all trades (informed or random, large or small) possibly contribute to long-term volatility. This corresponds to a paradigm change: instead of fundamental value determining prices, the main driver of price changes is the order flow itself --- whether informed or random, trades will move prices. This is the \textit{order-driven view of markets}, that progressively emerged in the last 20 years, motivated by several empirical facts obtained using microstructural data \cite{Madhavan,Hasbrouck,Lyons,Bouchaud04,Farmer05,Hopman,BFL,Deuskar,TQP}, to be recalled below.

At least naively, the order-driven theory of price changes offers a solution to the excess-volatility puzzle: if trades {\it by themselves} move prices, then excess trading could create excess volatility (see section \ref{sec:micro}). 
It is also likely to be the mechanism for the universal trend-following effect mentioned above, and would allow one to understand why self-exciting feedback effects are so prevalent in financial markets, leading to bursts of volatility even in the absence of any news \cite{Cutler,Joulin,Fosset,Marcaccioli}. In fact, calibrating -- say -- self-exciting Hawkes models on data leads to the conclusion that a very large fraction ($\gtrsim 80 \%$!) of market activity and \textit{volatility is self-generated}, rather than exogenous \cite{Sornette, Hardiman, Hawkes_review}.

Ironically, although appealing to professional traders, quantitative hedge funds (like CFM) and the layperson observing the stock market, the order-driven scenario where trade flows impact prices is viewed as ``sadly illiterate'' by most financial economists. As Rich Lyons noted in his seminal book \cite{Lyons}: {\it Consider an example that clarifies how economist and practitioner worldviews differ. The example is the timeworn reasoning used by practitioners to account for price movements. In the case of a price increase, practitioners will assert ``there were more buyers than sellers''. Like other economists, I smile when I hear this. I smile because in my mind the expression is tantamount to ``price had to rise to balance demand and supply''.} 

Yet the situation may start to shift in the wake of the rather awesome recent paper by Xavier Gabaix and Ralph Koijen (GK), entitled ``In Search of the Origins of Financial Fluctuations: The Inelastic Markets Hypothesis'' \cite{GK}. Based both on an empirical analysis of the long term price response to funds' order flow and on an equilibrium model of the holdings of mandate-constrained investment firms, the authors argue quite convincingly in favour of the order-driven scenario and debunk many dissenting arguments based on the traditional lore. The central result of GK is that buying (or selling) $1\$$ of an individual stock on average increases (decreases) the market capitalisation of that stock by $M \$$ in the long run, with $M \approx 1$ \textit{even for uninformed trades}. Buying the market as a whole (i.e. the index or a basket of stocks) has an even larger impact, with $M \approx 5$. The multiplier $M$ is therefore very substantial, when rational models would predict that uninformed trades should move the price only very mildly (i.e. $M \approx 0.01$, see \cite{GK}), if at all.

The GK story is based on global equilibrium considerations, as their ``undergraduate example'' illustrates: When an investor sells one dollar of bonds to invest in a representative asset manager with a mandate of keeping 80 \% of its investments in stocks, the price of stocks {\it has to increase}, in equilibrium, by five dollars, corresponding to $M=(1-80 \%)^{-1}=5$. However, the actual mechanism through which the price adjusts is left unspecified. This is where market microstructure comes into the picture. The aim of this paper is to discuss how the price adjustment actually unfolds, and why recent measures of {\it price impact} at the daily time scale, when correctly interpreted, are \textit{quantitatively} compatible with GK's value of the multiplier $M$. We believe that our reconciliation of high frequency liquidity (i.e. the realm of microstructure) and low frequency equilibrium (i.e. quarterly or longer, as studied by GK) is quite remarkable and gives strong credence to the order-driven/inelastic theory of markets.   

\section{High Frequency Price Impact}

\subsection{Two Sides to the Coin}\label{sec:2sides}

Buy trades (i.e. market orders hitting the ask) tend to push the price up and sell trades (i.e. market orders hitting the bid) tend to push the price down. This is price impact, and is an all-too-familiar reality for traders who need to buy or sell large quantities of an asset. To these traders, price impact is tantamount to a cost, because the impact of their earlier trades makes the price of their subsequent trades worse on average. 

In much of the existing literature, there are two strands of interpretation for price impact, which reflect the great divide between the efficient-market story and the order-driven story \cite{TQP}. At the two extremes of this spectrum are the following scenarios:

\begin{enumerate}
\item
{\it Agents successfully forecast short-term price movements, and trade accordingly.} This is the {efficient-market point of view} (see e.g. \cite{Hasbrouck}), which asserts that a trader who believes that the price is likely to rise will buy in anticipation of this price move, as in Kyle's model \cite{Kyle}. In this framework, a noise-induced trade that is based on no information at all can only have a short term impact on prices --- otherwise, prices would not behave as nearly perfect random walks as they do, and would end up straying very far from their fundamental values. 

By this interpretation, if the price was meant to move due to information, it would do so even {\it without} any trades.

\item {\it Price impact is a reaction to order-flow imbalance.} This is the {order-driven view}, which asserts that even if a trade reflected no information in any reasonable sense, then price impact would still occur and contribute to the long-term volatility. 
\end{enumerate}

In the first story, trades reveal private information about the fundamental value, creating a so-called {price discovery} process. In the second story, the act of trading itself impacts the price. In this case, one should remain agnostic about the information content of the trades, and should therefore speak of {price formation} rather than price discovery. But if market participants believe that the newly established price is the ``right'' price and act accordingly by revising their reservation price, ``information revelation'' might simply be a self-fulfilling prophecy. In section \ref{sec:micro}, we will discuss several empirical results which suggest that the amount of information per trade is actually very small, strongly favouring the order-driven scenario. 

\subsection{The Square-Root Law} 

In fact, traders or trading algos do not execute large trades via single market orders, but instead split up their trades into many small pieces. These pieces are executed incrementally, using market orders, limit orders, or both, over a period of several minutes to several days. The collection of all such individual orders belonging to the same trading decision is usually called a {\it metaorder}. How much does a metaorder of total volume $Q$, executed over a period of duration $T$, affect the price {\it on average}? Naively, it might seem intuitive that the impact of a metaorder should scale linearly in $Q$. Perhaps surprisingly, empirical analysis reveals that in real markets, this scaling is not linear, but rather is approximately square-root. 

\subsubsection{Empirical Evidence}

Since the early 1980s, a vast array of empirical studies spanning both academia and industry have concluded that the impact of a metaorder scales approximately as the square-root of its size $Q$ --- see \cite{TQP} for a recent review and an extensive list of references.  This result is reported by studies of different markets (including equities, futures, FX, options, and even Bitcoin), during different epochs (including pre-2005, when liquidity was mostly provided by market-makers, and post-2005, when electronic markets became dominated by HFT), in different types of microstructure (including both small-tick stocks and large-tick stocks), and for market participants that use different underlying trading strategies (including fundamental, technical, and so on) and different execution styles (including using mostly market orders, a mix of limit orders and market orders, or mostly limit orders, as in \cite{AQR}).

In all of these cases, the average (relative) price change between the beginning and the end of a metaorder with volume $Q$ is well-described by the ``square-root law'':
\begin{equation}\label{eq:square_root_law}
\mathfrak{I}(Q,T) \approx Y \sigma_{T} \sqrt{\frac{Q}{V_{T}}}, \qquad (Q \ll V_T)
\end{equation}
where $Y$ is a numerical coefficient of order 1 ($Y \approx 0.5$ for US stocks), $\sigma_{T}$ is the  contemporaneous volatility on the time horizon $T$, and $V_{T}$ is the  contemporaneous volume traded over time $T$. The square-root law of metaorder impact is well-established empirically, but there are several features that make it extremely surprising theoretically, at least at first sight. Before we discuss these surprising aspects, let us note that as with any empirical law, the square-root impact law holds (approximately) in the regime of intermediate execution horizons $T$ (not too fast nor too slow) and small enough volume fractions $Q/V_T$, which is the regime usually adopted by investors and execution algos in normal trading conditions (see \cite{TQP} for more on this). Different behaviours should be expected outside of these regimes, although data to probe these regions is scarce, and the corresponding conclusions currently remain unclear.

\subsubsection{Two Surprising Features}

The first surprising feature of Equation (\ref{eq:square_root_law}) is that metaorder impact does not scale linearly with $Q$ --- or, said differently, that metaorder impact is not additive. Instead, one finds empirically that the second half of a metaorder impacts the price much less than the first half. But this can only be the case if there is some kind of {liquidity memory time} $T_\text{m}$, such that the influence of past trades cannot be neglected for $T \ll T_\text{m}$ and vanishes for $T \gg T_\text{m}$, when all memory of past trades is lost. One therefore expects that beyond $T_\text{m}$, impact must become linear in $Q$. This is indeed what one finds within a model describing the dynamics of liquidity, which reproduces the square-root law at short times and a linear impact law at longer time -- see section \ref{sec:llob} below. The memory time $T_\text{m}$ will turn out to be very important to weld together the high frequency regime described in this section and the low frequency regime considered by GK.  

The second surprising feature of Equation (\ref{eq:square_root_law}) is that $Q$ does not appear  as a fraction of the total market capitalization ${\mathcal{M}}$ of the asset (as might be naively anticipated) but instead as a fraction of the total volume $V_T$ traded during the execution time $T$. In current equities markets, ${\mathcal{M}}$ is typically about $200$ times larger than $V_T$ for $T=$1 day. Therefore, the impact of a metaorder is \textit{much larger} than if the ratio $Q/{V_T}$ in Eq. (\ref{eq:square_root_law}) was instead $Q/{\mathcal{M}}$.\footnote{In the eighties, the lore was that trading $2\%$ of the daily volume, i.e. $0.01 \%$ of the market cap, would maybe move the price by a totally negligible $0.01 \%$. The Black-Scholes Delta-hedge induced crash of October 1987 was a dour wake-up call that impact is in fact not a small effect. See e.g. Treynor, J. L. (1988). Portfolio Insurance and Market Volatility. Financial Analysts Journal, 44(6), 71-73.
} The square-root behaviour for $Q \ll V_T$ also substantially amplifies the impact of small metaorders: executing $1 \%$ of the daily volume moves the price (on average) by $\sqrt{1 \%} = 10 \%$ of its daily volatility, 10 times larger than the naive, linear estimate.

The main conclusion here is that even relatively small metaorders cause surprisingly large impact. In fact, the GK multiplier measured in that regime would be uncannily high: taking $T=1$ day, $\sigma_{T} = 2.5 \%$ (a typical value for single stocks, corresponding to an annual volatility of $\approx 40 \%$) and $Q= 1 \% V_T = 5 \, 10^{-5} {\mathcal{M}}$ leads to $\mathfrak{I}(Q,T) \approx 1.25 \, 10^{-3}$ or $M = 25$. However, since impact is proportional to $\sqrt{Q}$ and not $Q$, GK's multiplier is meaningless in the regime $T \ll T_\text{m}$. As we will discuss in section \ref{sec:fromItoM}, the square-root impact contribution is mostly \textit{transient}, while the permanent part (which survives for $T \gg T_\text{m}$) is linear and characterized by $M$ of order unity, as indeed found by GK.   

\subsubsection{Possible Measurement Biases}\label{sec:biases}

The impact of a metaorder $\mathfrak{I}(Q,T)$ can be affected by several artifacts and biases. One of the recurrent criticism is that metaorders are not exogenous, and possibly conditioned on trading signals (see e.g. section \ref{sec:early} below) and/or on the price moves during the execution interval $T$. However, several arguments suggest that although such effects may in some cases considerably affect the measured impact curve, most of the data analyzed in the literature can be trusted. For one thing, CFM's proprietary data allows one to eliminate many of these biases, since the strength of the trading signal is known and can be factored in the regression. The fact that our own estimates of the square-root law precisely matches the ones reported in the literature (e.g. \cite{Almgren,AQR}), or the one that we measured using the Ancerno database containing metaorders issued by long term investors \cite{Zarinelli,Bucci}, is a strong argument in favour of the validity of the square-root law.\footnote{As a further important argument, note that the square-root law does not only describe the impact of the fully executed metaorder of volume $Q$, but the whole impact path $\mathfrak{I}(q,t)$ during execution, as a function of the partially executed volume $q \in [0,Q]$ and $t \in [0,T]$ \cite{Moro,TQP}.} As noted above, the same square-root law has been reported for a large variety of financial markets, including option markets \cite{Toth_options} or Bitcoin \cite{Donier_bitcoin}, suggesting a universal underlying mechanism which we discuss in section \ref{sec:llob} below. We have actually shown that the short term impact of CFM's trades is indistinguishable from the trades of the rest of the market \cite{Toth_others}, or, for that matter, from purely random trades that were studied at CFM during a specifically designed experimental campaign in 2010-2011. 

\section{The Origin of the Square-Root Law}\label{sec:llob}

\subsection{Early Theories}\label{sec:early}

Since the mid-nineties, several stories have been proposed to account for the square-root impact law. The first attempt, due to the Barra Group \cite{Barra} and Grinold \& Kahn \cite{Grinold} argues that the square-root behaviour is a consequence of market-markers being compensated for their inventory risk (see also \cite{YCZ}). The reasoning is as follows. Assume that a metaorder of volume $Q$ is absorbed by market-makers who will need to slowly offload their position later on. The amplitude of a potentially adverse move of the price during this unwinding phase is of the order of $\sigma_T \sqrt{T_{\text{off}}/T}$, where $T_{\text{off}}$ is the time needed to offload an inventory of size $Q$. It is reasonable to assume that $T_{\text{off}}$ is proportional to $Q$ and inversely proportional to the trading rate of the market $V_T/T$, giving $T_{\text{off}}/T=Q/V_T$. If market-makers respond to the metaorder by moving the price in such a way that their profit is of the same order as the risk they take, then it would follow that $\mathfrak{I} = Y \sigma_T \sqrt{Q/V_T}$, as  found empirically. However, this story assumes no competition between market-makers. Indeed, inventory risk is diversifiable over time, and in the long run averages to zero. Charging an impact cost compensating for the inventory risk of each metaorder would lead to formidable profits and would necessarily attract competing liquidity providers, eventually leading to a $Y$ coefficient much smaller than 1, at variance with empirical results, for which $Y = O(1)$.

Another theory, proposed by Gabaix et al. \cite{Gabaix_sqrt}, ascribes the square-root impact law to an information revelation effect, i.e. to the fact that trades are conditioned by some short-term predictability. One can show that in the presence of a short-term signal and linear impact, the optimal execution horizon $T^\star$ for metaorders of size $Q$ grows like $T^\star \sim \sqrt{Q}$ (see \cite{TQP}, section 21.2.3 for a detailed derivation). This theory argues that since the price is expected to move in the direction of the trade during $T^\star$ (as information is revealed), the peak impact itself behaves as $\sqrt{Q}$. However, this scenario would imply that the impact path is linear in the executed quantity $q$, whereas the full impact path in fact also behaves as a square-root of $q$ \cite{Moro,TQP}. Furthermore, most metaorders -- in particular those of CFM -- correspond at best to long term predictions, with very little short term ``alpha'' that would move the price of times scales $T \lesssim 1$ day.  

More recently, Farmer et al. \cite{FLW} proposed yet another theory that is very reminiscent of the Glosten--Milgrom model, which argues that the size of the bid--ask spread is actually set competitively. The theory assumes that  metaorders arrive sequentially, with a volume $Q$ distributed according to a power-law. Market-makers attempt to guess whether the metaorder will continue or stop at the next time step, and set the price such that it is a martingale and such that the average execution price compensates for the information contained in the metaorder (this condition is sometimes called ``fair-pricing''). If the distribution of metaorder sizes behaves precisely as $Q^{-5/2}$, these two conditions lead to a square-root impact law. Although enticing, this theory has difficulty explaining why the square-root impact law appears to be much more universal than the distribution of the size of metaorders or of the autocorrelation of trade signs. For example, the square-root law holds very precisely in 2013 Bitcoin markets, where the distribution of metaorder sizes behaves as $Q^{-2}$, rather than $Q^{-5/2}$, and at a time where market making was much less competitive \cite{Donier_bitcoin}. 

\subsection{The Latent Liquidity Theory}

Finally, T\'oth et al. \cite{Toth_PRX} proposed an alternative theory based on a dynamical description of supply and demand, called the Latent Liquidity Theory (LLT). This approach, developed further in several papers \cite{Mastro1,Mastro2,LLOB,Donier_Walras,Benzaquen1,Benzaquen2,QF}, provides a natural statistical interpretation for the square-root law and its apparent universality. It also makes further predictions, in particular concerning the {\it decay} of impact once the metaorder has been executed \cite{LLOB,Benzaquen1}. This is a piece of information that turns out to be crucial for recovering GK's multiplier at long times, about which the theories recalled above make substantially different predictions compared to LLT.  

A full discussion of the Latent Liquidity Theory is much beyond the scope of the present paper, and the reader is referred to \cite{TQP} for a detailed account and the Appendix for a short summary of its mathematical formulation. In a nutshell, the theory assumes that each long term investor in the market has a reservation price (to buy or to sell) that he or she updates as a function of time, due to incoming news, price changes, noise, etc. The collection of all these trading intentions constitutes the available liquidity at any instant of time -- although most of it remains ``latent'', i.e. is not immediately posted in the public order book. When the market price hits the reservation price of a given buy (sell) investor, his/her order is executed and becomes a sell (buy) intention, but at a price significantly higher (lower) than the execution price. (The theory does not consider HFT liquidity, which cannot provide much resistance to metaorders with an execution time $T$ longer than a few minutes). 

Reservation prices remain sticky during a typical memory time $T_\text{m}$ and, when revised, have a tendency to distribute themselves around the new, updated market price (see \cite{LLOB}, Appendix for a detailed discussion of this point). This assumption means that investors tend to mix their own price estimate with the market price, which is by itself a source of information about what other investors believe. Such an assumption looks very reasonable. Following Black's intuition \cite{Black}, fundamental value is so vaguely known (only up to a factor two says Black\footnote{\textit{We might define an efficient market as one in which price is within a factor of 2 of value, i.e.,  the price is more than half of value and less than twice value. The factor of 2 is arbitrary, of course. Intuitively, though, it seems reasonable to me, in the light of sources of uncertainty about value and the strength of the forces tending to cause price to return to value.} In his introduction, Black also writes: \textit{I recognize that most researchers [...] will regard many of my conclusions as wrong, or untestable, or unsupported by existing evidence. [...]. In the end, my response to the skepticism of others is to make a prediction: someday, these conclusions will be widely accepted. The influence of noise traders will become apparent.}}) that no one can really claim to know better than others, so some realignment of beliefs around the market price must take place (statistically speaking, of course).  

These are the main ingredients underpinning the LLT. On short time scales $T \ll T_\text{m}$, liquidity is essentially static and creates barriers to price motion. But as belief realignment happens, memory of previous intentions is erased and the impact of past trades on the price becomes \textit{permanent}. Embedding these ideas into a mathematical framework allows one to make precise predictions \cite{LLOB,TQP}, which can be summarized as follows:
\begin{enumerate}
    \item The impact of a metaorder of volume $Q$ executed over some time $T \ll T_\text{m}$ is given by the square-root law, Eq. (\ref{eq:square_root_law}). 
    \item Once the metaorder is fully executed, its impact decreases with time from its peak value
    Eq. (\ref{eq:square_root_law}) as a power-law of time, down to a value that depends on the information content that triggered the trade.
    \item In the absence of any information, the long-term, permanent part of the impact was explicitly computed in \cite{Benzaquen1} and is given by 
    \begin{equation}\label{eq:I_infty}
        I_\infty(Q,T) = \frac12 \sigma_1 \sqrt{T_\text{m}} \times \frac{Q}{V_1 T_\text{m}},
    \end{equation}
    where the subscript ``$1$'' corresponds to 1 day and $T_\text{m}$ is measured in days. Note that this result is in fact {\it independent} of the execution time $T$, which can be large or small compared to $T_\text{m}$. A brief summary of the calculation made in \cite{Benzaquen1} is provided in the Appendix, where the memory time $T_\text{m}$ is related to the latent liquidity renewal rate $\nu=1/T_\text{m}$.
\end{enumerate}

Hence, the permanent impact $I_\infty(Q,T)$ is {\it linear} in $Q$, independent of $T$, and proportional to the volatility on the scale of the memory time $T_\text{m}$, and inversely proportional to the volume traded by the market on the same time scale, $V_1 T_\text{m}$. A simple way to understand the result of 
\cite{Benzaquen1} is to notice that the $1/\sqrt{t}$ decay of impact, predicted in \cite{LLOB} and recalled in the Appendix as Eq. \eqref{priceeq_asympt}, must be interrupted after a time $\sim T_\text{m}$, since any memory of the initial price is erased beyond that time. This immediately yields Eq. \eqref{eq:I_infty}, up to a numerical factor.
  
In other words, whereas the square-root impact law describes the short term, transient response of the market to order-flow, the long-term, permanent component is much smaller, and linear in $Q$, in agreement with Almgren et al.'s early insight \cite{Almgren}. Note that the longer the memory time $T_\text{m}$, the smaller the permanent impact. This is in line with our LLT story above: the longer market participants stick to their beliefs, the stronger the anchor to a reference price and the smaller the long term impact. Of course, if liquidity was infinite (i.e. $V_1 = \infty$), or if beliefs were permanent (i.e. $T_\text{m}=\infty$) then markets would be perfectly elastic and transactions would not impact prices.

Finally, let us make an important remark: a very nice feature of the LLT is that any round trip incurs a positive cost, see \cite{LLOB} for a proof. To wit, trading impact prices, but there is no way to construct an arbitrage strategy out of this effect, since getting out of position would lead to a net loss.  

\section{The GK multiplier} \label{sec:fromItoM}

We now claim that it is the long term impact, Eq. (\ref{eq:I_infty}), that must be compared with GK's multiplier $M$. As we said above, it would not make sense to rely on Eq. (\ref{eq:square_root_law}) since (a) it is not linear in $Q$ and (b) it only describes the transient part of the impact, which will all but vanish on the quarterly time scale considered by GK \cite{GK}. 

More precisely, assuming $Q = 1 \% \mathcal{M}$ (i.e. a total order size of $1 \%$ of the market capitalisation, but any other number would do), the long term impact on the price expressed in percent is GK's multiplier, hence
 \begin{equation}\label{eq:I_infty_2}
        \boxed{M = \frac12 \sigma_{1} \frac{\mathcal{M}}{V_{1}} \times \sqrt{\frac{1}{T_\text{m}}}.}
\end{equation}
This is the central result of the present paper. Numerically, for $\sigma_{1}= 2.5 \%$ and ${\mathcal{M}}/{V_{1}}=200$, one finds $M=5/\sqrt{T_\text{m}}$. As noted above, the longer investors believe in their initial estimate of value and are ready to provide liquidity at that price, the more elastic the market and the smaller the GK multiplier.

The precise value of the memory time is difficult to pin down, since in fact we expect that market participants are characterized by a broad distribution of frequencies (see Appendix). But it is plausible that an effective value of $T_\text{m}$ is between a few days and a few weeks, see the data analyzed in Refs. \cite{Brokmann,Bucci_slow}. So choosing $T_\text{m}=20$ (corresponding to a month of trading) seems reasonable and yields $M \approx 1$, as obtained by GK. But note that the innocent looking result $M=O(1)$ turns out to be highly non-trivial, since it results from a large factor (${\mathcal{M}}/{V_{1}} \sim 100$) being compensated by a small factor ($\sigma_{1} \sim 1 \%$)! 

Our result furthermore makes falsifiable predictions. For example, assuming that $T_\text{m}$ is independent of the considered stock, we predict that the multiplier $M$ is proportional to $\sigma_{1}{\mathcal{M}}/V_{1}$. So one could expect that large capitalisation stocks, for which volatility is smaller and the fraction of market caps exchanged daily is larger, have a relatively smaller multiplier -- i.e., in the language of GK, larger capitalisation stocks are less ``inelastic''. 

\begin{figure*}
  \centering
  \includegraphics[width=0.75\linewidth]{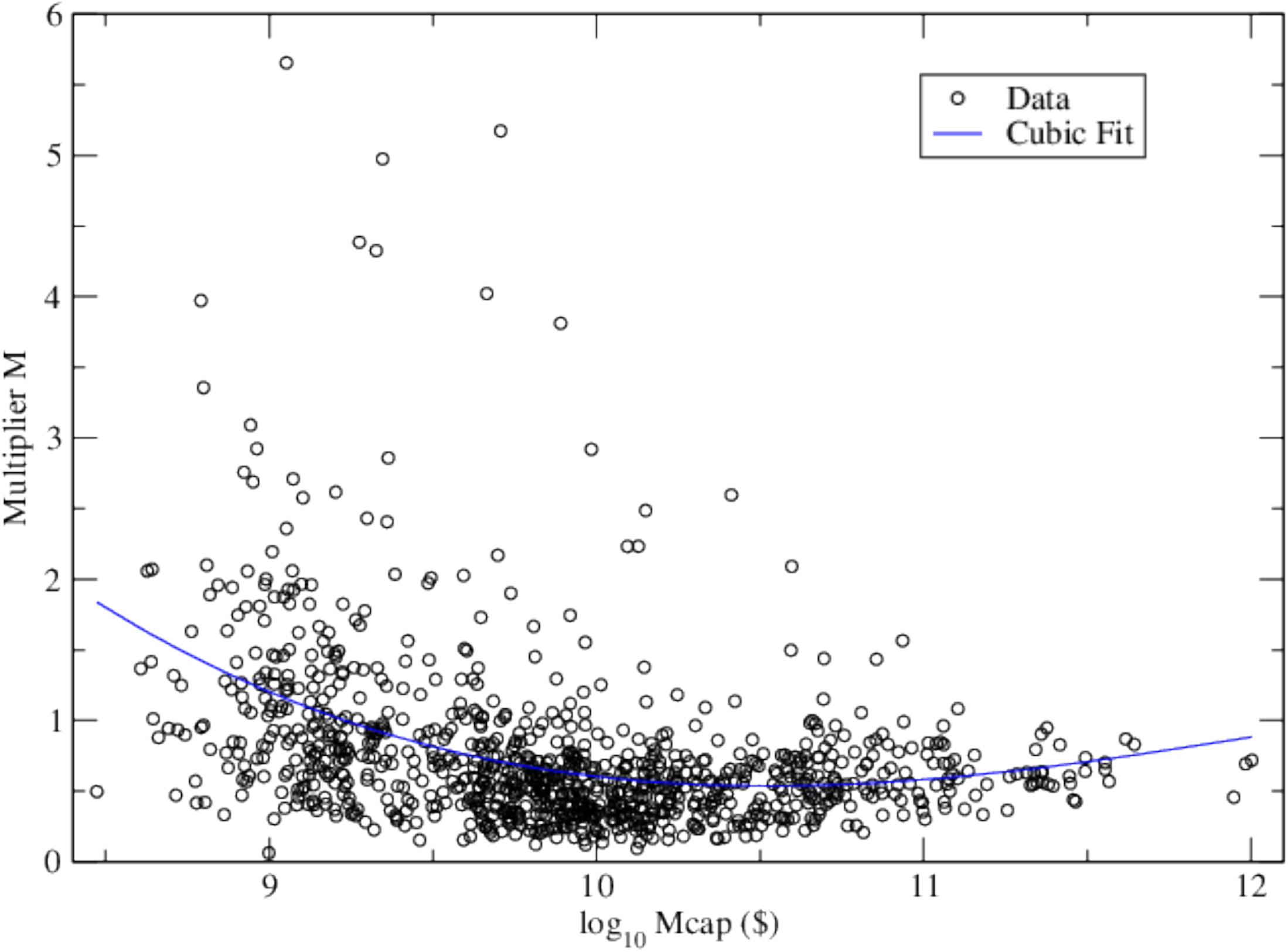}
  \caption{GK's multiplier $M$ as predicted by Eq. \eqref{eq:I_infty_3} (with an equal sign) for 950 US stocks in 2019 and with $\Delta = 10\%$. Volatility, average daily volume and market capitalisation are averaged over all days in 2019. With such a choice of $\Delta$, the average value of $M$ is found to be $0.76$ with a standard deviation of $0.6$. The plain line is a cubic fit as a function of $\log_{10} \mathcal{M}$, revealing a slightly non-monotonic variation of $M$ as a function of the market cap, perhaps increasing again for very large market caps. Note that the plot obtained using Eq. \eqref{eq:I_infty_2} with $T_\text{m}=16$ instead of  Eq. \eqref{eq:I_infty_3} looks very similar.}
  \label{fig}
\end{figure*}

But another plausible specification is to posit that the memory time $T_\text{m}$ is such that the volatility over that time scale reaches a certain universal threshold value $\Delta$, meaning that investors tend to realign their beliefs when the price level has changed ``appreciably'', say when $\Delta = 10 \%$ (corresponding to $T_{\text {m}}=16$ days for a $2.5 \%$ daily volatility). This is tantamount to setting $T_\text{m} = \Delta^2/\sigma_{1}^2$, which finally leads to\footnote{Note that Eq. \eqref{eq:I_infty_3} is the result one would get from a purely dimensional analysis if one assumes that there is no particular time scale in the problem.}
\begin{equation}\label{eq:I_infty_3}
        M \propto \frac{\sigma_{1}^2}{\Delta} \frac{\mathcal{M}}{V_{1}},
\end{equation}
where $\Delta$ is now assumed to be stock independent. It would be interesting to test this prediction more quantitatively using GK's methodology. However, as revealed in Fig. \ref{fig}, the overall range of variation predicted by Eq. \eqref{eq:I_infty_3} turns out to be rather restricted. The multiplier $M$ is seen to decrease by a factor roughly equal to $2$ between small cap stocks ($\mathcal{M} \sim 100M \$$) and large cap stocks ($\mathcal{M} \sim 1000B \$$), i.e. when $\mathcal{M}$ is multiplied by $10,000$. Still, with $\Delta = 10\%$, the average value of $M$ is found to be $0.76$ with a standard deviation of $0.6$, which means that the full span of variations of $M$ is up to a factor 10, which is consistent with the data shown in Ref. \cite{ESG}.

\section{Discussion}
\label{sec:micro}
\subsection{Statistical Efficiency vs. Fundamental Efficiency}

As is well known, asset prices are approximately martingales over time scales spanning from seconds to weeks. Another way to state this empirical fact is to say that volatility is approximately independent of the scale at which it is measured.  
In the efficient market picture, this phenomenon is a consequence of market prices only reacting to unpredictable news, and almost immediately digesting the corresponding information content. 

In practice, this explanation is hard to believe, because the fundamental value of an asset is only known so vaguely (as Black noticed, see above) that at least some amount of short-term mispricing should be present, even in very liquid markets. But this should induce {excess short-term volatility} and mean-reversion. For example, to be compatible with observations on the S\&P 500 futures contract, mispricings must be less than $0.05\%$ of the asset's price and have a reversion time of only 10 minutes. How can prices be so precise when there is so much uncertainty? 

In fact, the long-range autocorrelation in order flows \cite{BFL,TQP} is clear proof of the presence of \textit{long-lived imbalances} between supply and demand, which markets cannot immediately digest and equilibrate (as assumed in the efficient market picture) \cite{BFL}. Naively, these long-lived imbalances should create trends and mispricings. However, as argued in \cite{Bouchaud05,Wyart,TQP}, these effects are mitigated by liquidity providers who, in normal market conditions, compete to remove any exploitable price pattern and thereby buffer these imbalances. This is essentially the content of the so-called ``propagator model'' \cite{Bouchaud04}, in which impact decay is fine-tuned to compensate the long memory of order flow, and causes the price to be close to a martingale (see also \cite{Lillo_perm}). This makes prices \textit{statistically efficient} without necessarily being \textit{fundamentally efficient}. In other words, competition at high frequencies is enough to whiten the time series of returns, but not necessarily to ensure that prices reflect fundamental values. 

\subsection{The Long Term Fate of GK's Multiplier}

Within the order-driven view of markets, high-frequency traders and market makers only seek to exploit short term statistical arbitrage opportunities, without any long term view about fundamental value. By doing so, such traders activity makes prices unpredictable and simply propagate the high-frequency value of volatility to long time scales. The resulting volatility has no reason whatsoever to match the fundamental volatility. Hence, one plausible explanation for the excess-volatility puzzle is that the trading-induced volatility is much larger than the fundamental volatility. It is only over very long time scales (several years) that some mean reversion around the fundamental value can be observed, as surmised by Black \cite{Black} and substantiated in \cite{Black_was_right,Majewski,Suisse}. Correspondingly, we conjecture that the very long term ($> 5$ years) value of GK's multiplier $M$ is significantly smaller than the one measured on monthly time scales. Unfortunately, this long term limit will probably be very difficult to measure. 

\subsection{Volatility Equals Spread} 

As argued above, no-arbitrage at high frequencies is secured by HFT/market-making activities. But since these activities are highly competitive, one expects that the average profitability of liquidity provision is in fact close to zero. As argued in \cite{Wyart}, this condition is enough to enforce that spread and volatility are related. A simple framework to understand this relation is the MRR model \cite{MRR}, which is a bare-bone version of the propagator model \cite{Bouchaud04}. The upshot of the model is that the volatility {\it per trade} $\upsilon$ is given by \cite{Wyart,TQP}
\begin{equation} \label{MRR}
    \upsilon^2 =  \frac{1 - c_1^2}{4} s^2 +  \upsilon_0^2,
\end{equation}
where $s$ is the spread, $c_1$ is the one-lag auto-correlation coefficient of the sign of the trades and $\upsilon_0^2$ is the news induced contribution to volatility, i.e. price changes that would occur without trades. The usual (per unit time) volatility is then obtained as $\sigma_T^2 = \upsilon^2 N_T$, where $N_T$ is the average number of trades during time $T$.  

It turns out that Eq. \eqref{MRR} is remarkably obeyed by empirical data (see e.g. Fig. 16.6 in \cite{TQP}), with $\upsilon_0 \ll \upsilon$, meaning that the lion's share of the volatility is induced by trades, that is very little news induced jumps (on this point, see \cite{Joulin,Marcaccioli}). The very same conclusion is obtained within a more sophisticated version of the propagator model, that accounts for the non-markovian nature of the order flow (see e.g. \cite{TQP}, Chapter 14). 

The conclusion is that order flow, whether informed or non informed, is the major source of volatility in financial markets. This is of course also the content of the inelastic market story of GK: trades move the price, and the long term effect, measured by GK's multiplier $M$, is very substantial --- again, trading 1\% of the market cap moves the price by 1\%.  

In fact, we can provide another enticing interpretation of Eq. \eqref{eq:I_infty}, based on the relation \eqref{MRR} between spread and volatility:
\begin{equation}
    I_\infty \propto  \frac{s \, N_Q}{\sqrt{N_\text{m}}},
\end{equation}
where $s$ is the spread, $N_Q$ the number of individual trades needed to complete the execution of the metaorder, and $N_\text{m}$ the total number of trades taking place within the memory time interval $[0,T_\text{m}]$. Naively, each trade impacts the price by an amount proportional to the spread, but most of this impact decays as a power-law because of the autocorrelation of the sign of the trades, so on long timescales only a fraction $1/\sqrt{N_\text{m}}$ of the initial impact survives in the long run.

\subsection{Why Do Uninformed Trades Impact Prices To Start With?}

We wish to end this section with a short discussion of the standard paradox raised by the very notion of market impact. Since every buy trade is matched by a sell trades, why do trades impact prices at all, except if these trades anticipate some information which is only revealed later? This is the efficient market conundrum (see again section \ref{sec:2sides}) carefully resolved in GK's paper, where the idea of mandate-constrained asset managers crucially comes into play. 

From a microstructural point of view, each trade can be characterized as ``active'' (consuming liquidity) or ``passive'' (providing liquidity). This distinction breaks the symmetry at high frequencies, and allows one to define a meaningful order imbalance, as the sum of active buy trades minus the sum of active sell trades. It is this imbalance that reflects an aggregate ``urge'' to buy or sell and that mechanically impacts prices, whether or not this urge is justified by a genuine piece of information about future value. Ultimately, the GK multiplier $M$ will turn out to be the low frequency stigma of the asymmetry between active and passive trades.   

\subsection{GK's Multiplier For Stock Indices}

GK also argue that the multiplier $M$ is 5 times larger for the market as a whole (i.e. when buying the index) \cite{GK}. In their story, this comes from the fact that investors willing to sell stocks and substitute them with bonds are more scarce than investors willing to substitute one stock with another. Within a microstructure point of view, the amplification factor comes from ``cross-impact'', i.e. the fact of buying one stock pushes the price of all correlated stocks by a small, but measurable amount \cite{Benzaquen3}. Intriguingly, it turns out that this small cross-impact, when multiplied by the number of stocks in the index, recovers precisely the factor 5 suggested by GK (see \cite{Benzaquen3}, their Fig. 8 and the discussion thereafter). We again find this agreement quite remarkable, as it bolsters our claim that the mechanism underlying the inelastic hypothesis has a natural microstructural
origin. Note however that the LLT has not yet been generalized to account for cross-impact, which is an important open problem. 

\subsection{GK's Multiplier For Futures Markets}
\label{sec:futures}

Whereas GK's story chiefly concerns stock markets, our microstructural interpretation suggests a much broader applicability, in particular to commodity futures or foreign exchange. The only subtlety is to define the analogue of the market cap to obtain an a-dimensional ``multiplier''. One possibility is to use open interest, but it is not clear that such a choice is always meaningful. In any case, the LLT predicts that the medium term impact of buying one contract of any tradable asset is given by Eq. \eqref{eq:I_infty}, with $Q=1$ and $V_1$ measured in number contracts traded daily.

\section{Conclusion}

The aim of this paper was to relate Gabaix and Koijen's Inelastic Market Hypothesis \cite{GK} to the order-driven view of markets that emerged within the microstructure literature in the past 20 years. We reviewed the most salient empirical facts and arguments that give credence to the idea that market price fluctuations are mostly due to order flow, whether informed or non-informed: trades impact prices, even on the long run. We focused in particular on the Latent Liquidity Theory (LLT) of price impact, and argued that the underlying mechanism for what GK call inelasticity is the dynamics of private estimates of asset value, which tend to realign around the market price over some finite memory time that we called $T_\text{m}$. LLT in fact makes a precise prediction for GK's multiplier $M$, which measures the long term impact of transactions: if a trading firm buys $X \$$ of a company, the market capitalisation of that company increases by $M X \$$. Our central result is given by Eq. \eqref{eq:I_infty_2}, which relates $M$ to the daily volatility of the asset, the fraction of its market capitalisation that is traded daily, and the memory time $T_\text{m}$. Although trades permanently impact prices, there is no possibility of ``mechanical'' arbitrage within LLT.

The macro-finance implications of the inelastic market hypothesis are important and have been thoroughly discussed in GK's paper \cite{GK}, in particular the idea that governments could buoy the stock market by investing in equities \cite{Farmer}. From our point of view, the order-driven view of markets allows one to understand many of the puzzles of asset pricing theory, in particular the excess-volatility puzzle and the existence of long-lived bubbles and market rallies, fueled by a continuous inflow of buy orders, with the recent episode of Reddit meme stocks as a case in point \cite{Jaunin}. Note that the LLT theory is not restricted to stock, and predicts that similar effects also hold for any traded asset, see section \ref{sec:futures} above.

If order flow is the dominant cause of price changes, ``information'' is chiefly about correctly anticipating the behaviour of others, as Keynes envisioned long ago, and not about fundamental value. The notion of information should then be replaced by the notion of \textit{correlation with future returns}, induced by future flows. For example, when all market participants interpret a positive piece of news as negative and sell accordingly, the correct move for an arbitrageur is to interpret the news as negative, even if doing so does not make economic sense. Of course, if all market participants are rational and make trading decisions based on their best guess of the fundamental value, order flow will just reflect deviations from fundamentals and the efficient market picture is recovered. 

The idea that it is the order-flow that must be predicted, even if uninformed, resonates well with the intuition of finance professionals and allows one to understand why statistical regularities might exist and be exploited by quant firms. Indeed, flow data is quite popular among statistical arbitrage funds. The order-driven paradigm also allows one to resolve some paradoxes, like for example that it is surprisingly easier to find predictive signals for large cap. stocks than for small cap. stocks, probably because the former are more actively traded and that the order flow reveals more statistical regularities. The 2007 quant crunch and other recurrent deleveraging spirals are also extreme consequences of the impact of order flow on prices \cite{Lo_2007,Spirals,KO}. 

In conclusion, we hope that the present reformulation of the Inelastic Market Hypothesis in terms of mechanistic and measurable microstructural effects will shed a complementary light on the origin of financial market fluctuations, and possibly hammer a final nail into the coffin of the Efficient Market Hypothesis.

\subsection*{Acknowledgments}

I want to warmly thank M. Benzaquen (with whom Eq. \eqref{eq:I_infty} was derived), X. Gabaix, R. Koijen, I. Mastromatteo, D. Thesmar, B. T\'oth and Ph. van der Beck for many discussions around these specific topics, and Y. Lemp\'eri\`ere for providing the data used in Fig. \ref{fig}. Many of the ideas expressed in this paper were originally formulated in our book \cite{TQP} and I want to thank J. Bonart, J. Donier and M. Gould for a terrific collaboration.

\section*{Appendix: Permanent Impact within LLT}

We here briefly recall the main ingredients of the LLT as presented in \cite{LLOB}, see also \cite{TQP}. In the continuous limit we define the latent volume densities of limit orders in the order book as: $\varphi_{\mathrm{b}}(x,t)$ (buy) and $\varphi_{\mathrm{s}}(x,t)$ (sell). The latter evolve according to the following set of partial differential equations:
\begin{subeqnarray}
\partial_t \varphi_{\mathrm{b}} &=& \sigma_1^2 \partial_{xx}\varphi_{\mathrm{b}} -\nu\varphi_{\mathrm{b}} + \lambda \Theta(x_t-x) - R(x) \slabel{goveqsnl1}\\
\partial_t \varphi_{\mathrm{s}} &=& \sigma_1^2 \partial_{xx}\varphi_{\mathrm{s}} -\nu\varphi_{\mathrm{s}}  + \lambda \Theta(x-x_t) - R(x)\  ,\quad \ 
\slabel{goveqsnl2}
\end{subeqnarray}
where the different contributions on the right hand side respectively represent (from left to right): small random changes of agents' reservation prices (diffusion terms), cancellations with rate $\nu$ (death terms), arrivals of new intentions with intensity $\lambda$ (deposition terms), and finally matching of buy/sell intentions $R$ (reaction terms). 
The cancellation of orders corresponds to memory erasure and realignment of intentions around the current price, so the rate $\nu$ corresponds to the inverse of the memory time $T_\text{m}$ considered in the present paper:
\[
\nu = \frac{1}{T_\text{m}}.
\]

In the limit where $R \to \infty$ (corresponding to continuous double auction markets), buy and sell intentions cannot coexist and the market price $x_t$ therefore obeys $ \varphi_{\mathrm{b}}(x_t,t)= \varphi_{\mathrm{s}}(x_t,t) = 0$.
A crucial remark is that in that limit $\phi(x,t) = \varphi_{\textrm b}(x, t) - \varphi_{\textrm s}(x, t)$ solves a \textit{linear} equation \cite{Mastro2}:
\begin{eqnarray}
\partial_t \phi = \sigma_1^2 \partial_{xx} \phi -\nu\phi + s(x,t) \ ,\label{firsteqsrc}
\end{eqnarray}
where the deposition term reads $s(x,t) = \lambda \,\sgn (x_t-x)$. The stationary order book was computed by Donier \emph{et al.} \cite{LLOB} as: $\phi^\mathrm{st}(x)=-({\lambda}/{\nu}) \, \sgn(x)  [1-\exp(-p|x|) ]$
where $p=\sqrt{\nu/\sigma_1^2}$ denotes the typical length scale below which the order book can be considered as linear: $\phi^\mathrm{st}(x)=-\mathcal L x$ where $\mathcal L = \lambda/\sqrt{\nu \sigma_1^2}$ is a measure of liquidity, related to the volume traded per unit time $V_1$ through $V_1 = \sigma_1^2 \mathcal L$. 

Donier \emph{et al.} \cite{LLOB} focused on the \emph{infinite memory} linear order book limit, namely $\nu , \lambda \rightarrow 0$ (while keeping the liquidity $\mathcal L \sim \lambda {\nu}^{-1/2}$ constant), for which the impact of a metaorder asymptotically decays to zero, because agents never forget their initial beliefs. In \cite{Benzaquen1}, we have extended the calculation to small but non-zero $\nu$ (i.e. long memory time), for which some residual long-term impact is expected.  

The general solution of Eq.~\eqref{firsteqsrc} is given by:
\begin{eqnarray}
\phi(x,t) = \left( \mathcal G_\nu * \phi_0\right)(x,t) + \int \text d y\int_0^\infty \text d \tau\, \mathcal G_\nu(x-y,t-\tau) s(y,\tau) \ , \label{convol}
\end{eqnarray}
where $\phi_0(x) =\phi(x,0)$ denotes the initial condition, and 
\begin{eqnarray}
\mathcal G_\nu(x,t) = \max(t,0)  \frac{\exp\left[-\frac{x^2}{4\sigma_1^2 t}- \nu t\right]}{\sqrt{4\pi \sigma_1^2 t} } \ .
\end{eqnarray}
Following Donier \emph{et al.} \cite{LLOB}, we introduce a buy (sell) meta-order as an extra point-like source of buy (sell) particles with intensity rate $m=Q/T$, where $Q$ is the volume of the metaorder and $T$ the execution time, such that the source term in Eq.~\eqref{firsteqsrc} becomes: $s(x,t) = m \delta(x-x_t)\cdot \mathds{1}_{[0,T]} +\lambda \,\sgn (x_t-x)$. 

Performing the integral over space in Eq.~\eqref{convol} and setting $\phi_0(x)=\phi^{\mathrm{st}}(x)$ yields:
\begin{eqnarray}
\phi(x,t) =  \phi^\mathrm{st}(x)e^{-\nu t} + m \int_0^{t\wedge T} \text d \tau\,  \mathcal G_\nu(x-x_\tau,t-\tau)    -\lambda\int_0^{t } \text d \tau \,  \erf\left[  \frac{x-x_\tau}{\sqrt{4D(t-\tau)}}  \right] e^{-\nu(t-\tau)} \ .\label{mastereq}
\end{eqnarray}
The price $x_t$ solves the integral equation:
\begin{eqnarray}
\phi(x_t,t) = 0 \ .  \label{priceeq}
\end{eqnarray}
For $\lambda,\nu \to 0$ and for $t > T$, one immediately recovers Eq. (16) of \cite{LLOB}:
\begin{eqnarray}
x_t = x_t^0 = \frac{m}{\mathcal{L}} \int_0^{T} \text d \tau\,  \mathcal G_0(x_t-x_\tau,t-\tau),  \label{priceeq_asympt}
\end{eqnarray}
which boils down, at large $t$, to 
\begin{eqnarray}
x_t^0 \approx \frac{Q}{\mathcal{L}} \frac{1}{\sqrt{4\pi \sigma_1^2 t}} = \frac{\sigma_1}{\sqrt{4 \pi t}}  \frac{Q}{V_1}. \label{priceeq_asympt2}
\end{eqnarray}
Setting $t=T_{\text{m}}$ in this equation immediately leads to Eq. \eqref{eq:I_infty}, up to a numerical prefactor. 

In order to compute the long term impact exactly, the main idea of the calculation is to expand the price trajectory $x_t$ in powers of $\sqrt{\nu}$, i.e.
\begin{eqnarray}
x_t = x_t^0+\sqrt{\nu} x_t^1+O(\nu),\label{alphaz0z1}
\end{eqnarray}
where $x_t^0$ and $x_t^1$ respectively denote the 0th order and 1st order contributions. In the limit of short execution times ($T \ll T_\text{m}$) and small meta-order volumes $Q \ll V_\text{m}$, where $V_\text{m} = V_1 T_\text{m}$ is the total volume traded during the memory time $T_\text{m}$, one can look for a solution of the form $x^1_t= F(\nu t)$. In the long time limit $t\gg T$, using the zero-th order solution Eq. \ref{priceeq_asympt2} and setting $u = \nu t$, Eq.~\eqref{mastereq} boils down to  
\begin{eqnarray}
0= F(u) + \beta \int_0^{u } \text d v \,    \frac{\sqrt{v} -\sqrt{u}}{\sqrt{\pi uv(u-v)}}  e^{v} +  \int_0^{u } \text d v \,    \frac{F(u)-F(v) }{\sqrt{\pi(u-v)}}  e^{v}
\ ,\label{intequ}
\end{eqnarray}
where $\beta$ depends on the fast/slow nature of the execution (see \cite{Benzaquen1} for more details). The solution of this equation for $u \gg 1$ can is found to be 
\begin{eqnarray}
F(u)=F_\infty -\frac{\beta}{\sqrt{u}}\left[1-e^{-u}\right] \ ,\label{}
\end{eqnarray}
where $F_\infty = \frac12 \sigma_1 Q/V_1$ is completely independent of the trading speed. Since $x_t^0$ tends to zero at long times, the long term impact is given by the asymptotic value of $x_1^t = \sqrt{\nu} F_\infty$, which is the result given in Eq. \eqref{eq:I_infty}. Simply stated, Eq. \eqref{eq:I_infty} means that the long-term impact is a fraction of the price volatility over time scale $T_\text{m}$, where this fraction is given by the ratio of the volume of the metaorder $Q$ to the total volume traded on the same time scale $V_1T_\text{m}$. We believe that this intuitive result should be valid much beyond the specific set of hypotheses on which the above calculation is based.

As it is further discussed in \cite{Benzaquen1}, the assumption of a single memory time $T_\text{m}$ is not realistic, since investors with very different trading horizons co-exist in the market. If instead one assumes a distribution of memory times $\varrho(T_\text{m})$, the long-time impact is rather given by \cite{Benzaquen1}:
 \begin{equation}\label{eq:I_infty_4}
        I_\infty(Q) = \frac12 \sigma_1 \frac{Q}{V_1} \int_0^\infty {\rm d}x \, \frac{\varrho(x)}{\sqrt{x}},
    \end{equation}
where memory times $x$ are expressed in days, like the volatility $\sigma_1$ and the average daily volume $V_1$.

\end{document}